\documentclass[aps,prl,twocolumn,superscriptaddress,floats]{revtex4}
\usepackage{lineno}
\usepackage{amssymb}
\usepackage{graphicx}
\usepackage{natbib}
\setcitestyle{numbers,square}
\usepackage{amsmath}
\usepackage{amssymb}
\usepackage{graphicx}
\usepackage{CJK}
\usepackage{keyval}
\usepackage{dcolumn}
\usepackage{bm}
\usepackage{color}
\usepackage{txfonts}
\usepackage{verbatim}
\usepackage[rightcaption]{sidecap}

\usepackage{soul}
\usepackage{ulem}
\usepackage{graphicx}
\usepackage{epstopdf}
\usepackage{epsfig}
\usepackage{ulem}
\usepackage{float}
\usepackage{gensymb}

\begin{document}     

\title{High Temperature Superconductivity Dominated by Inner Underdoped CuO$_2$ Planes in Quadruple-Layer Cuprate (Cu,C)Ba$_2$Ca$_3$Cu$_4$O$_{11+\delta}$ }

\author{Xingtian Sun}
\affiliation{
Laboratory of Advanced Materials, State Key Laboratory of Surface Physics,
and Department of Physics, Fudan University, Shanghai 200438, China
}%

\author{Suppanut Sangphet}%
\affiliation{
Laboratory of Advanced Materials, State Key Laboratory of Surface Physics,
and Department of Physics, Fudan University, Shanghai 200438, China
}%

\author{Nan Guo}%
\affiliation{
Laboratory of Advanced Materials, State Key Laboratory of Surface Physics,
and Department of Physics, Fudan University, Shanghai 200438, China
}%

\author{Yu Fan}%
\affiliation{
Laboratory of Advanced Materials, State Key Laboratory of Surface Physics,
and Department of Physics, Fudan University, Shanghai 200438, China
}%

\author{Yutong Chen}%
\affiliation{
Laboratory of Advanced Materials, State Key Laboratory of Surface Physics,
and Department of Physics, Fudan University, Shanghai 200438, China
}%

\author{Minyinan Lei}%
\affiliation{
Laboratory of Advanced Materials, State Key Laboratory of Surface Physics,
and Department of Physics, Fudan University, Shanghai 200438, China
}%

\author{Xue Ming}%
\affiliation{
Center for Superconducting Physics and Materials National Laboratory of Solid State Microstructures and Department of Physics, Nanjing University, Nanjing 210093, China.}

\author{Xiyu Zhu}%
\affiliation{
Center for Superconducting Physics and Materials National Laboratory of Solid State Microstructures and Department of Physics, Nanjing University, Nanjing 210093, China.}

\author{Hai-Hu Wen}%
\email{hhwen@nju.edu.cn}
\affiliation{
Center for Superconducting Physics and Materials National Laboratory of Solid State Microstructures and Department of Physics, Nanjing University, Nanjing 210093, China.}
\affiliation{
Collaborative Innovation Center of Advanced Microstructures, Nanjing University, Nanjing 210093, China.
}%

\author{Haichao Xu}
\email{xuhaichao@fudan.edu.cn}

\author{Rui Peng}
\email{pengrui@fudan.edu.cn}
\affiliation{
Laboratory of Advanced Materials, State Key Laboratory of Surface Physics,
and Department of Physics, Fudan University, Shanghai 200438, China
}%

\author{Donglai Feng}
\email{dlfeng@hfnl.cn}
\affiliation{New Cornerstone Science Laboratory, Hefei National Laboratory, Hefei, 230026, China}

\date{\today}

\begin{abstract}

The superconducting transition temperature ($T_{\mathrm{c}}$) of trilayer or quadruple-layer cuprates typically surpasses that of single-layer or bilayer systems. This observation is often interpreted within the ``composite picture", 
where strong proximity effect between inner CuO$_2$ planes (IPs) and outer CuO$_2$ planes (OPs) is crucial.
Albeit intriguing, a straightforward scrutinization of this composite picture is still lacking.
In this study, using angle-resolved photoemission spectroscopy to investigate (Cu,C)Ba$_2$Ca$_3$Cu$_4$O$_{11+\delta}$ (CuC-1234) with a high $T_{\mathrm{c}}$ of 110~K, we found that the OPs are not superconducting at the $T_{\mathrm{c}}$ of the material. Instead, the large pairing strength and phase coherence concurrently emerge at the underdoped IPs, suggesting that the high $T_{\mathrm{c}}$ is primarily driven by these underdoped IPs. Given that the $T_{\mathrm{c}}$ of CuC-1234 is comparable to other trilayer or quadruple-layer cuprates, our findings suggest that the conventional “composite picture” is not universally required for achieving high $T_{\mathrm{c}}$.
More importantly, we demonstrate that CuO$_2$ planes free of apical oxygen can support superconductivity up to 110~K even at a doping level of 0.07 holes per Cu, a level that lies deep in the underdoped regime of single- and bilayer cuprates.
These findings provide new insights into the origin of high $T_{\mathrm{c}}$ in multilayer cuprates.

\end{abstract}
\maketitle

\textit{Introduction}---In cuprate superconductors, the highest superconducting transition temperature ($T_{\mathrm{c}}$) is consistently observed in tri-layer or quadruple-layer systems \cite{Tc_vs_n(1),Tc_vs_n(2),CuC1234_sum}. Understanding the mechanism behind the high $T_{\mathrm{c}}$ in multilayer cuprates may hold the key ingredients to design new high-temperature superconductors and further enhance $T_{\mathrm{c}}$. To understand this, the unique structural feature in trilayer or quadruple-layer cuprates should be considered: the outer CuO$_2$ planes (OPs), located adjacent to the charge reservoir layers (CRLs), possess overdoped hole concentration, while the inner CuO$_2$ planes (IPs), sandwiched between two OPs, maintain underdoped carrier concentration and preserve a flat, square-planar CuO$_2$ structure \cite{Bi2223_multilayer,Bi2223dopinglevel,ZhouXingjiang_Bi2223, OP_IP_doping, OP_IP_doping2}. However, it remains an intriguing issue how these planes interact and why the coexistence of these two distinct doping levels and structural characteristics in the CuO$_2$ planes leads to such a high $T_{\mathrm{c}}$.

Several scenarios have been proposed. A widely discussed scenario, known as the ``composite picture,'' emphasizes the proximity effect and interlayer hopping ($t_{\perp}$) between OPs and IPs. In this model, IPs provide strong pairing but lack phase coherence due to low carrier density, while OPs offer phase coherence owing to higher doping but rely on the IPs for pairing. Their coupling allows both layers to become superconducting at the same $T_{\mathrm{c}}$, effectively combining their advantages to support a higher $T_{\mathrm{c}}$.
\cite{proximity_in_multilayer, Composite_picture, composite_detail1, Composite_OP,ZhouXingjiang_Bi2223,Bi2223_multilayer}. However, an alternative perspective emphasizes the intrinsic properties of IPs, suggesting that their cleaner, flatter CuO$_2$ planes with homogeneous and appropriate hole doping are the key drivers of high $T_{\mathrm{c}}$, while the OPs play a supporting role by protecting the IPs, minimizing defect densities, and modulating doping levels \cite{IPclean1, IPclean2,EsakiPRB2004}.
Although both scenarios are intriguing, they lack direct experimental verification.
One crucial test is to suppress superconductivity in either the OPs or IPs to disentangle the roles of the OPs and IPs.
Here, we report that, at the superconducting $T_{\mathrm{c}}$ of 110~K of the single-crystal quadruple-layer cuprate (Cu,C)Ba$_2$Ca$_3$Cu$_4$O$_{11+\delta}$ (CuC-1234), superconducting gap and coherence peaks are only observed at IPs, while the superconductivity of OPs develops below $T_{\mathrm{c2}}$=~70$\pm$5~K. Our findings indicate that pairing strength and phase stiffness originate predominantly from IPs, which primarily contribute to the high $T_{\mathrm{c}}$ superconductivity. These results indicate that ``composite picture'' is not universally required for achieving high $T_{\mathrm{c}}$. Instead, underdoped CuO$_2$ planes without apical oxygen can, under certain conditions, support superconductivity up to 110~K. These findings offer new perspectives on the mechanisms underlying high $T_{\mathrm{c}}$ in multilayer cuprates.

\begin{figure}[!htbp]
\includegraphics[width=86mm]{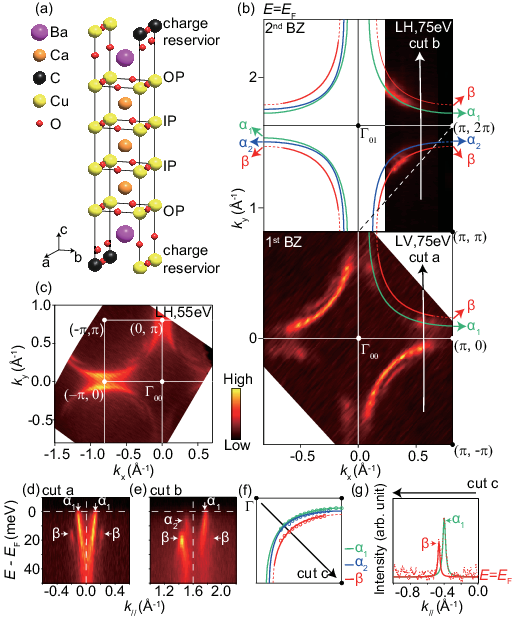}
\caption{Fermi surface topology. 
(a), Layered structure of CuC-1234 single crystal in a unit cell \cite{CuC1234_neutron}, which shows two outer CuO$_2$ planes (OPs) 
and two inner CuO$_2$ planes (IPs) away from the charge reservoir layer (CRL). 
(b), Photoemission intensity maps at the first Brillouin zone (1$^{st}$ BZ) and at the second Brillouin zone (2$^{nd}$ BZ). The red, green and blue curves indicate the three sets of Fermi surface sheets, $\beta$, $\alpha_1$ and $\alpha_2$, respectively. The dashed lines indicate the antiferromagnetic Brillouin zone boundaries.
(c), Photoemission intensity map showing higher intensity around (0, $\pi$). All maps are integrated over [$E_F$-10~meV, $E_F$+10~meV] and measured at T=8~K.
(d)-(e), Band dispersions of CuC-1234 measured along cuts a and b as indicated in (b).
(f), Superimposed Fermi surface sheets of $\beta$, $\alpha_1$ and $\alpha_2$. The open circles indicate the position of Fermi crossings. 
(g), The momentum distribution curve (MDC) of cut c (indicated in (f)) integrated at $E_F$. The red dots are raw data fitted by two Lorentz functions (red and green lines). The full width at half maximum (FWHM) of $\beta$ and $\alpha_1$ MDCs are 0.0285$\AA$$^{-1}$ and 0.0426$\AA$$^{-1}$, respectively.
}
\label{Fermi surface}
\end{figure}

\begin{figure*}[htbp]
\includegraphics[width=170mm]{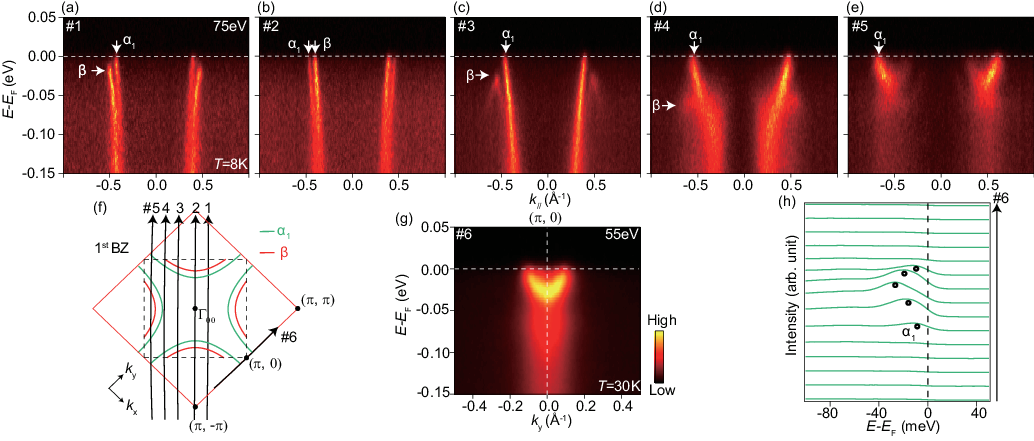}
\caption{Band dispersions at 1$^{st}$ BZ. 
(a)-(e), Photoemission intensity from nodal to antinodal regions [cuts~\#1-\#5 in panel (f)]. The white arrows indicate the gap or Fermi crossing of the $\beta$ and $\alpha_1$ bands. 
(f), Illustration of $\alpha_1$ and $\beta$ Fermi surface sheets in the 1$^{st}$ BZ. 
(g), Photoemission intensity measured along cut~\#6 in panel (f). 
(h), Energy distribution curves (EDCs) along cut~\#6. The black circles indicate the coherence peak positions of the EDCs.
}
\label{dispersion}
\end{figure*}

Within the (Cu,C)Ba$_2$Ca$_{n-1}$Cu$_n$O$_{2n+3}$ family, quadruple-layer CuC-1234 exhibits the highest $T_{\mathrm{c}}$  ($T_c\sim$116~K) \cite{CuC1234_sum,CuC1234_origin}, which is also comparable to the maximum $T_{\mathrm{c}}$  observed in other trilayer or quadruple-layer cuprate systems. The unit cell of CuC-1234 contains: (i) CRLs made up of CuO chains and CO$_3$ units, (ii) two equivalent OPs with pyramidal oxygen coordination and tilted apical oxygens, (iii) two equivalent IPs with square oxygen coordination [Fig.~1(a)].
Nevertheless, electronic structure studies on quadruple-layer cuprates with high-$T_{\mathrm{c}}$ ($\geq$77~K) have been a missing piece so far, largely due to historical challenges in synthesizing high-quality single crystals.
Recent success in crystal growth has enabled the fabrication of CuC-1234 single crystals with $T_{\mathrm{c}}$ of 110~K \cite{CuC1234_synthesized}, offering opportunities for angle-resolved photoemission spectroscopy (ARPES) studies. 

\textit{Fermi surfaces}---The photoemission intensity at Fermi energy ($E_{\mathrm{F}}$) [Fig.~1(b)] resolves two well-separated Fermi surface sheets (labeled $\alpha_1$ and $\beta$) in the first Brillouin zone (1$^{\mathrm{st}}$ BZ) at 8~K. In the second Brillouin zone (2$^{\mathrm{nd}}$ BZ), while both sheets persist, an additional Fermi surface sheet ($\alpha_2$) can be identified in the lower right part of the 2$^{\mathrm{nd}}$ BZ, whose shape deviates from that of $\alpha_1$. More obviously, the photoemission cut at the 2$^{\mathrm{nd}}$ BZ reveals asymmetric dispersion around ($\pi, 2\pi$)  [Fig.~1(e)], in marked contrast to the symmetric dispersion around antinode ($\pi, 0$) observed in the 1$^{\mathrm{st}}$ BZ cut [Fig.~1(d)], which further corroborates that the upper and lower regions of the 2$^{\mathrm{nd}}$ BZ originate from two distinct electronic bands, $\alpha_1$ and $\alpha_2$ bands. 
The selective visibility of $\alpha_1$ and $\alpha_2$  features at particular momenta can be attributed to the different photoemission matrix elements of bonding and anti-bonding bands. Based on the Luttinger theorem \cite{Luttinger1960}, the hole doping level of the $\beta$, $\alpha_1$ and $\alpha_2$ Fermi surface sheets are estimated to be 0.07~holes per Cu atom, 0.32~holes per Cu atom, and 0.25~holes per Cu atom, respectively [Fig.~1(f)]. Considering that IPs generally possess lower hole doping \cite{OP_IP_doping, OP_IP_doping2}, the $\beta$ sheet can be attributed to the IP-dominated Fermi surface. The $\alpha_1$ and $\beta$ sheets are separated in momenta spanning from nodal (($\pi/2$,$\pi/2$)) to antinodal regions, similar to the Fermi surfaces of IPs and OPs in Bi$_2$Sr$_2$Ca$_2$Cu$_3$O$_{10+\delta}$ (Bi2223) \cite{Bi2223dopinglevel,ZhouXingjiang_Bi2223}. The Fermi crossing of $\alpha_2$ pocket coincides with the $\alpha_1$ sheet in the near nodal region and increasingly deviates from $\alpha_1$ towards the antinode, similar to the bilayer splitting observed in Bi$_2$Sr$_2$CaCu$_2$O$_{8+\delta}$ (Bi2212) \cite{Feng_Bi2212_bilayer_splitting1, chuang2001_Bi2212_bilayer_splitting2, ai2019_Bi2212_bilayer_splitting3} and OPs in overdoped Bi2223 \cite{ZhouXingjiang_Bi2223}. The absence of bilayer splitting in more closely spaced IPs could be attributed to the significantly lower doping level and stronger electron correlations in the IPs \cite{IP_lowdoping_1,IP_lowdoping_2,IP_lowdoping_3}.
Compared to the general phase diagram of cuprate \cite{Diagram1}, the IPs are underdoped, while the OPs, with an average doping of 0.285 holes per Cu, are heavily overdoped. 
Along the nodal direction where the $\alpha_1$ and $\alpha_2$ bands degenerate, the $\beta$ band shows a sharper peak in the momentum distribution curve (MDC) than $\alpha_1$ band [Fig.~1(g)], suggesting lower scattering rates in the IPs \cite{MDC_scattering}.

\textit{Superconducting gap structure}---The spectral weight of $\beta$ Fermi surface is suppressed away from the nodal region [Fig.~1(b)], a signature of Fermi arc behavior, aligning with the behavior in other underdoped or optimally doped cuprates \cite{Bi2212_Fermiarc, underdoped_PG, underdoped_PG3, underdoped_Fermiarc2}. By contrast, the $\alpha_1$ Fermi surface shows strong spectral weight at $E_{\mathrm{F}}$ around antinode and forms a closed $\alpha_1$ sheet [Fig.~1(c)].
As shown by the photoemission spectra along a series of momenta at the 1$^{\mathrm{st}}$ BZ [Figs.~2(a)-2(e)], while the $\beta$ band displays a $d$-wave-like gap, \textit{i.e.} crossing $E_{\mathrm{F}}$ near the node and gradually gapping out toward the antinode, the $\alpha_1$ band consistently crosses $E_{\mathrm{F}}$ at all momenta. Especially, along cut $\#$6, $\alpha_1$ shows electron-like dispersion with clear Fermi drop, in stark contrast to the large gap of the OPs in other multilayer systems including Bi2223 \cite{ZhouXingjiang_Bi2223,Bi-based_ARPES, chen2006_F0234(1), chen2009_F0234(2)}.
This dispersion is similar to an ordinary metal, resembling Bi2201 in the normal state \cite{Bi2201AN2, Bi2201AN}. Such distinctions highlight the different behaviors in OPs between CuC-1234 and Bi2223.

In the 2$^{\mathrm{nd}}$ BZ, the superconducting gap structures of the three Fermi surface sheets, $\beta$, $\alpha_1$, and $\alpha_2$ are further examined [Fig.~3]. Consistent with the observation in the 1$^{\mathrm{st}}$ BZ, the $\beta$ band exhibits a $d$-wave gap following a linear dependence on $|\cos (k_{x}a) - \cos (k_{y}a)|/2$ [Figs.~3(f) and 3(i)], and the $\alpha_1$ band shows no sign of gap opening across the entire momentum range from nodal to antinodal regions [Figs.~3(b)-3(e), 3(h)]. For the $\alpha_2$ band, which is only observable in the 2$^{\mathrm{nd}}$ BZ, it exhibits a sizable superconducting gap around the antinode [Fig.~3(g)], with the gap size being smaller than that of the $\beta$ band. The larger superconducting gap in the bonding branch ($\alpha_2$) compared to the anti-bonding branch ($\alpha_1$) is consistent with observations in overdoped Bi2223 \cite{ZhouXingjiang_Bi2223}. Although the superconducting gap of the $\alpha_2$ band slightly increases from off-nodal to antinodal regions [Fig.~3(g)], no gap is observed near the nodal region. This could be attributed to a dirty $d$-wave gap function in $\alpha_2$ or an artifact from a gapless $\alpha_1$ band mixing into the spectra. The superconducting gap sizes of both the $\alpha_1$ and $\alpha_2$ bands are significantly smaller than those of the OPs in optimally doped Bi2223 [Fig.~3(i)] and overdoped Bi2223 \cite{ZhouXingjiang_Bi2223}, indicating weaker superconductivity in the OPs of CuC-1234. Intriguingly, the gap size of the $\beta$ band closely resembles that of the IPs in Bi2223 [Fig.~3(i)]. Notably, the doping level of the IPs in CuC-1234 (0.07 holes per Cu) matches that of the IPs in optimally doped and overdoped Bi2223 \cite{Bi2223dopinglevel,ZhouXingjiang_Bi2223}, and their $T_{\mathrm{c}}$'s are also identical. 

\begin{figure*}[htbp]
\includegraphics[width=170mm]{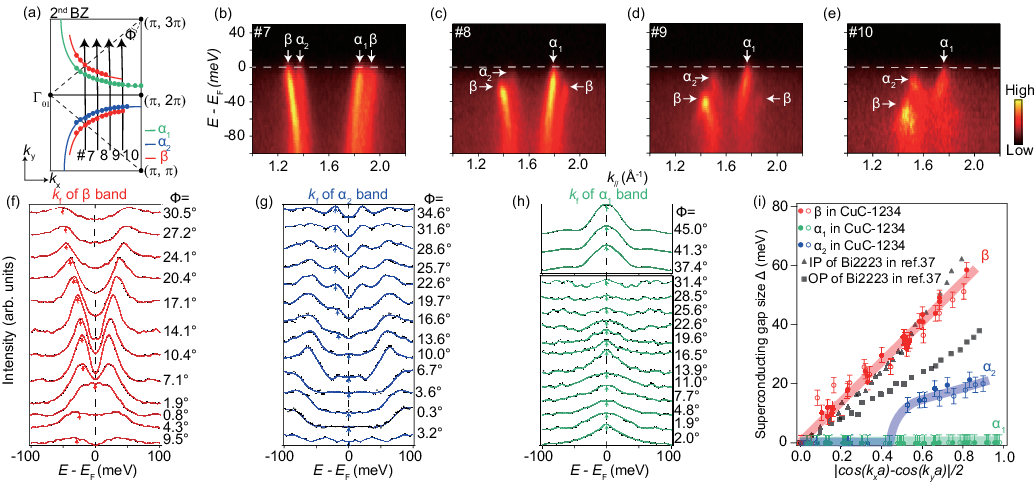}
\caption{Band evolution at 2$^{nd}$ BZ.
(a), Illustration of $\alpha_1$, $\alpha_2$, and $\beta$ Fermi surface sheets in the 2$^{nd}$ BZ.
(b)-(e), Photoemission intensity showing band dispersion of $\alpha_1$, $\alpha_2$, and $\beta$ from nodal to antinodal regions [cuts~\#7-\#10 in panel (a)], respectively. Precise momentum-dependent band dispersions of $\alpha_1$, $\alpha_2$, and $\beta$ bands are shown in extended data [see Supplemental Material, Fig.~S1].
(f)-(h), Symmetrized EDCs along the $k_{\mathrm{F}}$'s of $\beta$, $\alpha_2$ and $\alpha_1$, respectively. $\Phi$ indicates the polar angle of $k_{\mathrm{F}}$'s with respect to ($\pi$, $\pi$), as indicated in (a). The upper panel of (h) is the data from 1$^{st}$ BZ, while all other data are from 2$^{nd}$ BZ. The arrows track the superconducting coherence peaks.
(i), Superconducting gap sizes of $\beta$, $\alpha_1$, and $\alpha_2$ as a function of $|cos(k_xa)-cos(k_ya)|/2$, with the optimally doped Bi2223 also plotted for comparison \cite{Bi2223anticrossing}. 
The open circles are data extracted from panels (f)-(h), and filled circles are from extended data [see Supplemental Material, Fig.~S2]. 
All data of CuC-1234 were taken in the superconducting state at 8~K. The error bar is set to $\pm 2.6meV$, which is 20\% of the energy resolution 13~meV \cite{Resolution}.
}

\label{Gap evolution}
\end{figure*}

\textit{Temperature-dependence of superconducting gaps}---As the temperature increases, the superconducting gap of $\beta$ band around node closes at 110~K [Fig.~4(b)], and the spectral weight near $E_{\mathrm{F}}$ in the near-antinodal region dramatically decreases at 110~K [see Supplemental Material, Figs.~S3(b)-S3(c)], consistent with the $T_{\mathrm{c}}$ of CuC-1234 from transport measurements \cite{CuC1234_synthesized}. 
Moreover, in the near-antinodal region, a pseudogap emerges and persists above 130~K [see Supplemental Material, Fig.~S3(a)]. These observations can be interpreted by the IPs being in the underdoped regime \cite{underdoped_PG, underdoped_PG2, underdoped_PG3, underdoped_PG4}. 
In contrast, for the $\alpha_1$ band in the near-antinodal region, neither superconducting gap nor pseudogap is observed from 130~K down to 7~K like a normal metal [Fig.~4(c)]. 
For the $\alpha_2$ band,  the superconducting gap closes around 70~K [Figs.~4(d)-4(g), Fig.~4(h)], with the coherence peak disappearing as well [Fig.~4(i)]. These results suggest that while the $\beta$ band opens a superconducting gap and exhibits coherence peaks at the bulk $T_{\mathrm{c}}$=110~K,
$\alpha_2$ band has a lower critical temperature $T_{\mathrm{c2}}$=70~K $\pm$5~K [Figs.~4(h)-4(i)]. 
This observation corroborates earlier nuclear magnetic resonance (NMR) studies on polycrystalline CuC-1234, which revealed an anomaly in the Knight shift and spin-lattice relaxation rate around 60~K in addition to the change near the bulk $T_{\mathrm{c}}$, suggesting a secondary transition $T_{\mathrm{c2}}$ associated with the OPs \cite{CuC1234_NMR1, CuC1234_NMR2}.

\textit{Discussion}---The lower $T_{c2} = 70 \pm 5$~K of the $\alpha_2$ band, along with the absence of a superconducting gap in the $\alpha_1$ band down to 7~K, shows that the OPs are not superconducting at the bulk $T_{\mathrm{c}}$ of 110~K. Even in the absence of superconducting OPs, the IPs possess a large pairing gap and strong phase stiffness, and the system superconducts at a high $T_{\mathrm{c}}$ comparable to Bi2223.
These results imply that IPs play a dominant role in enabling the high $T_{\mathrm{c}}$ in CuC-1234. In addition, the distinct $T_{\mathrm{c}}$ from OPs, along with the BCS-like temperature dependence of the gap and coherence peak weight in the $\alpha_2$ band [Figs.~4(h)-4(i)], suggests that $T_{\mathrm{c2}}$ arises from intrinsic superconductivity in the OPs rather than being induced via proximity from the IPs.
Despite the negligible proximity effect between the OPs and IPs, the system still reaches a high $T_{\mathrm{c}}$ of 110~K, 
demonstrating that strong interlayer coupling is not a prerequisite for high-$T_{\mathrm{c}}$ superconductivity.
These characteristics suggest that the composite picture is not universally required for achieving high $T_{\mathrm{c}}$ in multilayer cuprates.

The observed superconducting transition around 70~K at $\alpha_2$ with $p \approx 0.25$, and the absence of superconductivity at $\alpha_1$ with $p \approx 0.32$, are roughly consistent with the phase diagram of Bi2212 \cite{Bi2212_phasediagram}. The corresponding CuO layers contain apical oxygens.
On the other hand, IPs of CuC-1234 can dominate superconductivity up to 110~K even at a doping level of 0.07 holes per Cu, a deeply underdoped regime of single- and bilayer cuprates. This likely suggests a different phase diagram for square planar CuO$_2$ planes free of apical oxygen. Rather than directly contributing to superfluid density as in the composite picture, OPs might play a supporting role. The enhancement of superconductivity in this OP-IP coexisting structure might stem from the following two aspects:

\begin{figure*}[htbp]
\includegraphics[width=170mm]{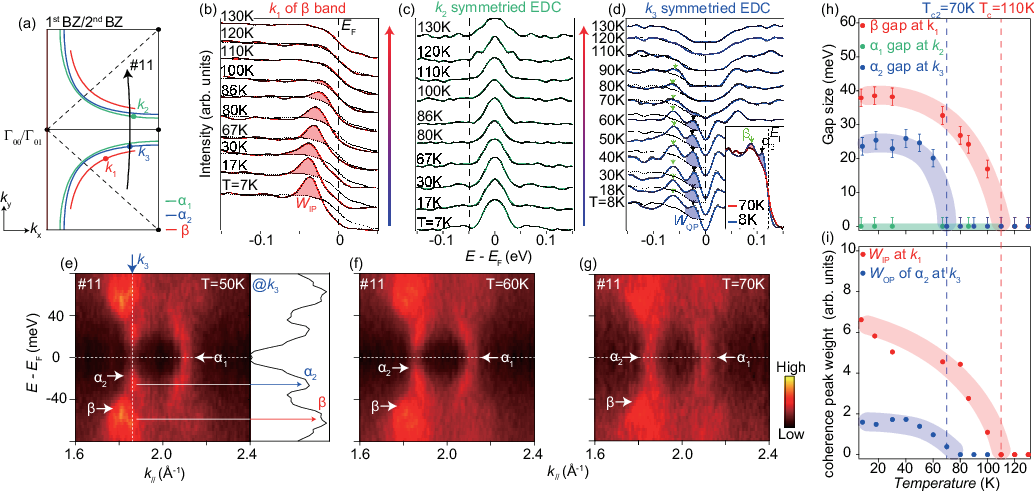}
\caption{Temperature dependent superconducting gap.
(a), Illustration of $\alpha_1$, $\alpha_2$, and $\beta$ Fermi surface sheets combining the 1$^{st}$ and the 2$^{nd}$ BZ.
(b)-(d), Temperature dependent EDCs at $k_{1}$ of $\beta$ band and symmetried EDCs at $k_{2}$ and $k_{3}$ of $\alpha_1$ and $\alpha_2$ bands, respectively.
The EDC of normal state ($T$=130~K) are superimposed at other EDC spectra in panels (b) and (d). The red and blue shaded areas represent the superconducting coherence peak weight of $\beta$ band ($W_{\mathrm{IP}}$) and $\alpha_2$ band ($W_{\mathrm{OP}}$), respectively. 
(e)-(g), Symmetrized photoemission intensities along cut \#11 in panel (a) taken at different temperatures. The symmetried EDC at $k_3$ is plotted on the right side of panel (d). 
(h), Superconducting gap size of $\beta$, $\alpha_1$ and $\alpha_2$ bands at $k_1$, $k_2$ and $k_3$ as functions of temperature, respectively. 
(i), Superconducting coherence peak weight $W_{\mathrm{IP}}$ at $k_1$ and $W_{\mathrm{OP}}$ at $k_3$ as functions of temperature, respectively. 
}
\label{T-dependent}
\end{figure*}

(i) \textbf{The unique environment of the IPs.} In multilayer cuprates, the IPs adopt a square-planar CuO$_2$ structure free of apical oxygen and are well separated from the charge reservoir layers by the OPs.
Apical oxygen is known to amplify disorder from the charge reservoir layers to the CuO$_2$ planes \cite{EsakiPRB2004}, and resonant inelastic X-ray scattering experiments have shown that it is detrimental to high-$T_{\mathrm{c}}$ superconductivity \cite{Apical_O1}.
The absence of apical oxygen in IPs therefore protects them from such disorder. Moreover, the OPs serve as a buffer layer, further shielding the IPs from structural and chemical disorder originating from the reservoir layers \cite{EsakiPRB2004}.
This protective environment enables more uniform hole doping and reduced lattice distortion in the IPs compared to single-layer and bilayer cuprates.
Experimentally, this is supported by the sharper momentum distribution curves (MDCs) of the $\beta$ band (from IPs) compared to the broader $\alpha_1$ band (from OPs), indicating that the IPs are cleaner and better suited to host robust superconductivity [Fig.~1(g)].

(ii) \textbf{Capacitive coupling between the IPs and OPs}. Prior theoretical and experimental studies have demonstrated that in a two-dimensional Josephson-Junction array with strong superconducting fluctuations, dissipation from an adjacent metallic layer can suppress phase fluctuations and thereby enhance superconductivity \cite{Capacitive1, Capacitive2}. 
Notably, unlike the composite picture requiring interlayer electron hopping, this capacitive coupling relies solely on junction capacitance.
In underdoped cuprates, strong superconducting fluctuations often lead to a loss of phase coherence \cite{Phase_fluctuation}, which may explain why a CuO$_2$ plane with 0.07 holes per Cu in single-layer or bilayer systems barely exhibits superconductivity. In contrast, in CuC-1234, superconducting fluctuations in the IPs can induce current fluctuations in the metallic OPs through the capacitive coupling. The fluctuating currents dissipate in the non-superconducting OPs, thereby suppressing the superconducting fluctuations in the IPs. This dissipation channel helps stabilize phase coherence and ultimately enhances superconductivity.

To summarize, our ARPES studies on the quadruple-layer cuprate CuC-1234 offer detailed insights into the Fermi surface topology and superconducting gap structure of both the IPs and OPs. Our findings challenge the universality of the ``composite picture", suggesting that the high $T_{\mathrm{c}}$ in CuC-1234 primarily originates from the underdoped IPs, while the non-superconducting  OP might play a supporting role above its own $T_{\mathrm{c2}}$. Remarkably, a square-planar CuO$_2$ plane without apical oxygens and with a doping level of 0.07 holes per Cu can sustain superconductivity up to 110~K. This is in contrast to the conventional phase diagram of cuprates with apical oxygens, where such low doping typically corresponds to a deeply underdoped, low-$T_{\mathrm{c}}$ state. These results provide new perspectives on the mechanisms driving high-$T_{\mathrm{c}}$ superconductivity in multilayer systems and offer new insights for enhancing $T_{\mathrm{c}}$ in cuprates.

\textit{Acknowledgements}---We gratefully acknowledge Dr. Zhengtai Liu, Dr. Timur. K. Kim and Dr. Chris Jozwiak for experimental support during the beamtime and Dunghai Lee for helpful discussions. We thank Shanghai Synchrotron Radiation Facility for access to beamline 03U, and the Advanced Light Source for access to beamline 7.0.2. This work is supported in part by the National Natural Science Foundation of China (grant Nos.12274085, 92477206, 12422404, 92365302), the National Key R$\&$D Program of the MOST of China (2023YFA1406300), the New Cornerstone Science Foundation, the Innovation Program for Quantum Science and Technology (Grant No.2021ZD0302803), and Shanghai Municipal Science and Technology Major Project (Grant No.2019SHZDZX01).

\newpage


\end{document}